---------------------------------------------------------------
\documentstyle[12pt,a4]{article}

\newcommand{\be}{\begin{equation}}
\newcommand{\ee}{\end{equation}}

\title{Projectivised representations of U$_q$osp(2,2)}

\author{Y. BRIHAYE\\
Department of Mathematical Physics\\
University of Mons\\
Av. Maistriau, B-7000 MONS, Belgium.\\
Stefan Giller{$^{*}$},
       Piotr Kosinski \thanks{$^{\dagger}$ Work supported by grant $n^o$
KBN 2P30221706p02}\\
Department of Theoretical Physics\\
University of Lodz\\
Pomorska 149/153, 90-236 Lodz, Poland}
\begin{document}

\begin{titlepage}
\maketitle
\thispagestyle{empty}
\begin{abstract}
We construct representations of the envelopping algebra U$_q$osp(2,2)
in terms of finite difference operators and we discuss this result in the
framework of the quasi-exactly solvable equations.
\end{abstract}
\vfill
\end{titlepage}

\section{Introduction}
One of the attractive features of quasi exactly solvable (Q.E.S.)
equations  \cite{label1,label2}
is that they provide nice interplays between spectral problems
 and abstract algebraic structures. Following
the approach of Refs. \cite{label1,label3,label4} it appears that the basic
ingredients needed for the construction (and the classification
\cite{label4}) of Q.E.S. equations are the projectivized representations
of some algebra, i.e. the representations formulated in terms of differential
operators of one or several variables. The most celebrated example is
the SL(2,$\reel$) algebra represented by the operators
\begin{equation}
J^+_n = x^2{d \over {d x}} - nx\quad ,\quad  J^0_n = x{d \over
{d x}}-{n\over 2}\quad , \quad J^-_n = {d \over  {d x}}
\end{equation}
which preserves the space of polynoms of degree $n$ in the variable $x$,
say $P_n$, if $n$ is a positive integer \cite{label4}. Elements of the
envelopping algebra generated by the operators (1) provide
as many differential operators that preserve $P_n$. They
can be transformed into hermitian operators
whose spectrum can be partly obtained by solving algebraic equations,
in this respect they are called quasi-exactly solvable operators.
\par Another example of interest is related to the super Lie algebra
osp(2,2) which possess projectivised representations formulated in
terms of 2$\times$2 matrix operators \cite{label3}
\begin{eqnarray}
T^{\epsilon} &= \left(\begin{array}{cc}
J^{\epsilon}_{n-1} &0\\
0                  &J^{\epsilon}_n \end{array} \right) \quad,
\quad
&J = -{1\over 2} \left(\begin{array}{cc}
n+1 &0\\
0 &n
\end{array}\right)\nonumber \\
\bar Q_1 &= (x{d\over dx}-n)\left(\begin{array}{cc}
0 &1\\
0 &0
\end{array}\right)\quad , \quad &\bar Q_2 = {d\over {dx}}
\left(\begin{array}{cc}
0 &1\\
0 &0
\end{array}\right)\\
Q_1 &= \left(\begin{array}{cc}
0 &0\\
1 &0
\end{array}\right) \quad , \quad &Q_2 = x \left(\begin{array}{cc}
0 &0\\
1 &0
\end{array}\right)\nonumber
\end{eqnarray}
where $\epsilon = \pm,0$.If $n$ is a positive integer, the operators (2)
preserv
$P_{n-1,n}$, the space of the couples of polynoms with degree $n-1$
(resp. $n$) for the first (resp. second) component.
\par Owing the actual interest for deformed
algebra in mathematical physics [5], it is natural to try to construct
the projectivised representations for deformed versions of the
algebras involved in  Q.E.S. problems; i.e. representations
formulated in terms of a Jackson symbol (see Eq.(4)).
The operators obtained in this way  provide the finite
difference counterparts to the usual Q.E.S. equations.
A projectivised representation of the algebra U$_q$SL(2,$\reel$)
is presented in \cite{label4}.
The generators read
\begin{eqnarray}
&\tilde j^+ = &q^{-n \over 2} (x^2 D_q - [n]x) \nonumber \\
&\tilde j^- = &q^{-n \over 2} D_q \\
&\tilde j^0 = & { 2\over q+1} {(n+1) q^{-n}  \over [n+1]}
                     (x D_q - {[n][n+1]\over [2n+2]}) \nonumber
\end{eqnarray}
with
 \begin{equation}
  [n]_q \equiv {1-q^n\over {1-q}} \quad ,\quad
D_q f \equiv  {f(x)-f(qx)\over {(1-q)x}}
\end{equation}
and fulfil the following quommutation rules
\begin{eqnarray}
q \tilde j^0 \tilde j^- - \tilde j^- \tilde j^0  &=& - \tilde j^-
\nonumber \\
q^2 \tilde j^+ \tilde j^- - \tilde j^- \tilde j^+ &=& -(q+1) \tilde j^0 \\
  \tilde j^0 \tilde j^+ - q \tilde j^+ \tilde j^0  &=& - \tilde j^+
\nonumber \end{eqnarray}
known as the second Witten's deformation
in the classification of Ref.\cite{label5}.
Standard formulas  allows one to transform the $\tilde j$
into new operators  whose (normal) commutators close
within their envelopping algebra \cite{label5}.

In this note, we aim to construct the counterpart of Eqs.(2)
for a deformation of the algebra osp(2,2).
Such deformations were first discussed in Ref.\cite{label6} and more recently
in Ref.\cite{label7}. After specifying   our  deformation of
osp(2,2) we will  present  a set of finite dimensional representations
of it.
The matrix elements of these reprentations can be
expressed in terms of a variable $x$
and of the finite difference operator $D_q$, providing
a projectivised representation of the algebra
under investigation.
The initial deformation is choosen in such a way that the
projectivised representations are simple and close to Eq.(2).

\par To specify  our version of  U$_q$osp(2,2) we
 use the notations of Ref.\cite{label6} : the bosonic generators
are denoted $H,J_{\pm},T$ and the fermionic ones by $V_{\pm}$ and
$\bar V_{\pm}$. The (anti-)commutation rules of the undeformed algebra
read as  follows:
\begin{equation}
[H,J_{\pm}] = \pm J_{\pm}\quad , \quad [J_{\pm},T]=0\quad , \quad
[H,T]=0
\end{equation}
\begin{equation}
[J_+,J_-] = 2H
\end{equation}
\begin{equation}
[H\pm {1\over 2} T, V_{\pm}] = 0 \quad ,\quad [H \mp {1\over 2} T,
V_{\pm}] = \pm V_{\pm}
\end{equation}
\begin{equation}
       [H \pm {1\over 2} T, \bar V_{\pm}] = \pm \bar V_{\pm}\quad ,
 \quad [H \mp {1\over 2} T, \bar V_{\pm}] = 0
\end{equation}
\begin{equation}
\lbrace V_i, V_j\rbrace = \lbrace \bar V_i, \bar V_j\rbrace = 0
\quad (i,j = \pm)
\end{equation}
\begin{equation}
\lbrace V_{\pm},\bar V_{\pm}\rbrace = \pm {1\over 2} J_{\pm}
\end{equation}
\begin{equation}
\lbrace V_+, \bar V_-\rbrace = -{1\over 2} (H+{1\over 2}T)\quad ,
\quad \lbrace \bar V_+, V_-\rbrace = - {1\over 2} (H-{1\over 2} T)
\end{equation}
Along with Ref.\cite{label6} we assume that the relations (8,9,10,11) are
preserved by the deformation which is introduced at the level of
Eq. (12) :
\begin{equation}
\lbrace V_+,\bar V_-\rbrace = {1\over 2} [P_+]_q \quad ,
\quad \lbrace \bar V_+,V_-\rbrace = -{1\over 2} [P_-]_q
\quad {\rm with} \quad
P_{\pm} \equiv H + {1\over 2} T\quad
\end{equation}
Adopting Eq. (11) as a definition of the operators $J_{\pm}$
and using the various Jacobi identities, one can
show that the relations (6) are preserved while others, involving
$J_{\pm}$, lead in general to deformed expressions, e.g.
\begin{eqnarray}
&[J_{\pm},V_{\mp}] &= V_{\pm} q^{P_{\mp}-{1\over 2} \pm {1\over 2}}\\
&[J_{\pm},\bar V_{\mp}] &= \bar V_{\pm} q^{P_{\pm}-{1\over 2} \pm {1\over
2}}\\
&[J_+,J_-] &= [P_+]_q q^{P_-} + [P_-]_q q^{P_+} + 2(1-{1\over q})
(\bar V_+V_-q^{P_+} + V_+\bar V_-P^{P_-})
\end{eqnarray}
The last relation indicates that $J_{\pm}$ and $H$ do not close
within their enveloping algebra for $q \neq 1$. It can be also verified that
the
following mapping $\Delta$ constitutes a coproduct on U$_q$osp(2,2).
\begin{equation}
\Delta H = H \otimes I + I\otimes H
\end{equation}
\begin{equation}
\Delta T = T \otimes I + I\otimes T
\end{equation}
\begin{equation}
\Delta V_{\pm} = I \otimes V_{\pm}  + V_{\pm} \otimes q^{P_{\pm}/2}
\end{equation}
\begin{equation}
\Delta \bar V_{\pm} = I \otimes \bar V_{\pm}  +
\bar V_{\pm} \otimes q^{P_{\mp}/2}
\end{equation}
In order to construct the representation of interest for us, we
assume that the space of the representation is the direct sum of two
subspaces respectively annihilated by $V_{\pm} $ and $\bar V_{\pm}$ :
\begin{equation}
H = H_d \oplus H_{u}  \quad , \quad
V_{\pm} H_{u} =0 \quad , \quad  \bar V_{\pm} H_d = 0
\end{equation}
and that the basic vectors are labelled according to their eigenvalues
with respect to the commuting operators $H$ and $T$.
\begin{equation}
T|h,u> = t_u|h,u> \quad , \quad T|\tilde h,d> = t_d|\tilde h,d>
\end{equation}
\begin{equation}
H|h,u> = h|h,u> \quad , \quad H|\tilde h,d> = \tilde h|\tilde h,d>
\end{equation}
After some algebra, one can show that the representation is finite
dimensional if
\begin{equation}
t_u = n \quad , \quad t_d = n+1\quad , \quad n\ \ {\rm integer}
\end{equation}
\begin{equation}
-{n\over 2} \leq h \leq {n\over 2}\quad , \quad -{m\over 2}
\leq \tilde h \leq {m\over 2}\quad ,\quad  n \equiv m+1
\end{equation}
with $h, \tilde h$ varying by unit steps.
The action of the other operators on the basic states then takes the form
\begin{equation}
J_{\pm}|h,u> = \mp [\mp {n\over 2} + h]_q |h\pm 1,u>
\end{equation}
\begin{equation}
J_{\pm}|\tilde h,u> = \mp [\mp {m\over 2} +\tilde h]_q |h\pm 1,d>
\end{equation}
\begin{equation}
\bar V_{\pm}|h,u> = [h\mp {n\over 2}]_q |h\pm {1\over 2},d>
\end{equation}
\begin{equation}
V_{\pm}|\tilde h,d> = -{1\over 2} |\tilde h \pm {1\over 2},u>
\end{equation}
In order to construct  the projectivised representation,
it is worth identifying the basic vectors
with some monomials in a suitable space of polynoms.
Let us pose
\begin{equation}
|h,u> \equiv \left(\begin{array}{c}
0\\ x^{h+{n\over 2}}
\end{array}\right)\quad , \quad
|\tilde h, d> = \left(\begin{array}{c}
x^{\tilde h+{m\over 2}} \\ 0 \end{array}\right)
\end{equation}
then we can  express the relations (22-29) in terms
of operator  $D_q$ and of the variable $x$ and
obtain a projectivised representation of the algebra
under investigation. The explicit form of the generators reads
\begin{equation}
J_- = \left(\begin{array}{cc}
D_q &0\\
0 &D_q\end{array}\right)\quad , \quad
J_+ = -x^2
\left(\begin{array}{cc}
x^mD_q x^{-m} &0\\
0 &x^nD_qx^{-n}
\end{array}\right)
\end{equation}
\begin{equation}
T  = \left(\begin{array}{cc}
n+1 &0\\
0 &n \end{array}\right)\quad , \quad q^H =
\left(\begin{array} {cc}
1+(q-1) x^{1+{m\over 2}}D_qx^{-{m\over 2}}
&0\\
0 & 1+(q-1)x^{1+{n\over 2}} D_q x^{-{n\over 2}}\end{array}\right)
\end{equation}
\begin{equation}
\bar V_+ = x^{n+1}D_qx^{-n}
\left(\begin{array}{cc}
0 &1\\
0 &0
\end{array}\right)\quad , \quad
\bar V_- = D_q \left(\begin{array}{cc}
0 &1\\
0 &0\end{array}\right)
\end{equation}
\begin{equation}
V_- =  -{1 \over 2} \left(\begin{array}{cc}
0 &0 \\
1 &0\end{array}\right)\quad , \quad
V_+ = -{x\over 2} \left(\begin{array}{cc}
0 &0\\
1 &0\end{array}\right)
\end{equation}

The set of operators (2) was used \cite{label8}
to classify two kinds of linear, differential equations having
polynomial solutions: the equations of two variables
(one real and one Grassmann) and the 2$\times$2 matrix
equations. The operators (2) and (31-34) share  the
property of preserving the space of couples of polynomials
with degrees $n-1$ and $n$, in this respect they can be used
to study and classify the counterparts of the two problems
above mentionned, once differential operators are replaced
by finite difference operators.
\vskip 2 cm

\noindent {\bf Acknowledgements.}\\
 We would like to thank Prof. J. Lukierski
          for pointing  Ref.\cite{label6} to our attention.

\end{document}